\documentclass{tPHM2e}

\usepackage{amssymb}
\usepackage{hyperref}
\usepackage{graphicx}
\usepackage{amsmath,amssymb,amsfonts}

\begin{document}

\title{Use of frit-disc crucibles for routine and exploratory solution growth of single crystalline samples}

\author{
\name{Paul C. Canfield$^{\ast}$\thanks{$^\ast$Corresponding author. Email: canfield@ameslab.gov} Tai Kong, Udhara. S. Kaluarachchi, and Na Hyun Jo}
\affil{Ames Laboratory and Department of Physics and Astronomy, Iowa State University, Ames, Iowa 50011, U.S.A.}}



\maketitle

\begin{abstract}
Solution growth of single crystals from high temperature solutions often involves the separation of residual solution from the grown crystals. For many growths of intermetallic compounds, this separation has historically been achieved with the use of plugs of silica wool. Whereas this is generally efficient in a mechanical sense, it leads to a significant contamination of the decanted liquid with silica fibers. In this paper we present a simple design for frit-disc alumina crucible sets that has made their use in the growth single crystals from high temperature solutions both simple and affordable. An alumina frit-disc allows for the clean separation of the residual liquid from the solid phase. This allows for the reuse of the decanted liquid, either for further growth of the same phase, or for subsequent growth of other, related phases. In this paper we provide examples of the growth of isotopically substituted TbCd$_{6}$ and icosahedral i-$R$Cd quasicrystals, as well as the separation of (i) the closely related Bi$_{2}$Rh$_{3}$S$_{2}$ and Bi$_{2}$Rh$_{3.5}$S$_{2}$ phases and (ii) PrZn$_{11}$ and Pr$_{2}$Zn$_{17}$.   
\end{abstract}



\section{Introduction}

Solution growth of single crystalline materials is arguably the most versatile and widely used method of producing bulk crystalline materials for basic and applied research. In the realm of condensed matter physics, and more specifically the studies of the electronic and magnetic properties of novel materials, solution growth has been the primary technique for the discovery and growth of materials with new or exotic properties and/or phase transitions. When considering a solution growth several key parameters have to be defined: number of elements in melt; composition of melt; temperature profile, e.g. maximum and minimum temperatures and ramp rate; and containment of melt.  It is this last point, how to contain the melt and, ultimately separate the residual liquid from the desired crystalline phase, that is the focus of this paper.

One of the key innovations in solution growth of single crystals was the routine use of a centrifuge to separate the hot, residual melt/solution from the crystalized phase at the end of growth\cite{1,2,3}. Simple laboratory centrifuges readily produce rotational frequencies of 2000 rpm and can have effective radii (sample to axis of rotation) of 15 cm; such a system can produce accelerations 600 times Earth's gravitational acceleration. This allows for the exceptionally efficient removal of residual solution from crystalline faces, often resulting in mirrored, as grown, surfaces\cite{4,5,6,7}.  

In order for this process to separate the remaining liquid from the crystalline solid phase some form of physical filter has to be used.  Originally, a plug of moderately compressed silica wool would be inserted into an inverted catch crucible that was placed on top of the growth crucible. This is described in references \cite{1,2,3} with specifically detailed pictures in reference \cite{3}. As more reactive melts were used, e.g. solutions with abundances of rare earth, or alkaline/alkaline earth elements, we had to develop a filter that could be made out of Ta or other materials. For the Ta-based crucibles that we use for the growth of $R$-Mg-Zn quasicrystals\cite{8}, Nd$_{2}$Fe$_{14}$B crystals\cite{6}, or exploration of Li- and Ca-based melts\cite{7}, we developed what is now referred to as a 3-cap crucible. For these growths Ta is used to avoid thermite-type reactions that would occur with an Al$_{2}$O$_{3}$ crucible. A length of Ta tube is cut and the top and bottom are ultimately sealed by welding in a Ta cap fabricated from Ta sheet. The growth and decant sides of this space are separated by a third Ta cap that has small holes drilled into it so as to allow it to serve the role of a frit or sieve. (See figure 1 of ref. \cite{7} for picture and details.)  When we used solutions that were incompatible with both Al$_{2}$O$_{3}$ as well as Ta we extended the incorporation of a drilled out frit to a BN based crucible set. In this case, we could actually machine the crucibles and drilled, frit-disc to have threads so as to allow the three of them to screw together\cite{9}. Based on this, we recently returned to Al$_{2}$O$_{3}$; taking advantage of machinable alumina we were able to develop crucibles for high-temperature growth (above the softening point of silica) that could be screwed together so as to form a single, closed assembly that separated the growth and decant (also called spin or catch) sides with a drilled frit-disc. (See reference \cite{9} for pictures.) 

Although the threaded alumina crucibles described in reference \cite{9} are required if the assembly has to be mechanically assembled into a single sealed object, the threading of the crucibles and drilled frit-disc is time consuming and expensive, making a three piece threaded set much more expensive than a simple pair of crucibles. In this paper we present a simplified, more versatile, and much less expensive variant of the frit-disc crucible set and illustrate how its routine use is far superior to the more traditional silica wool plug. We have found that use of a frit disc for separation of the crystals from the solvent, as compared to silica wool, allows for quantitative analysis of the yield as well as reuse and analysis of the uncontaminated decanted liquid.

\section{Technique Details}

\begin{figure}
\begin{center}
\includegraphics[scale = 0.8]{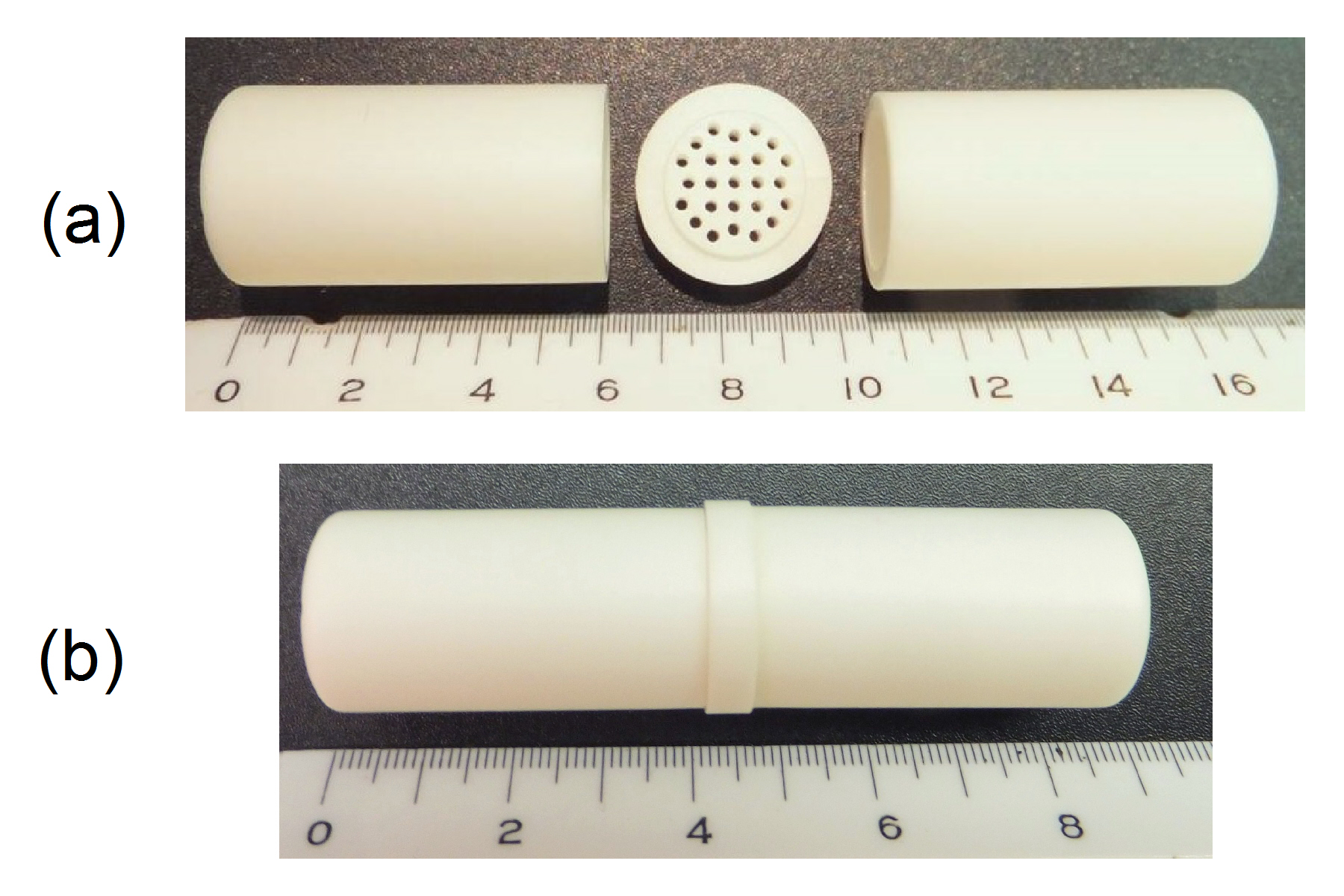}
\end{center}
\caption{2 ml frit-disc crucible set. (a) The frit-disc has a milled shoulder to make what can be considered to be a “step-taper” that allows for co-axial alignment of the three part assembly. (b) The assembled crucible set\cite{10}.}
\label{1}
\end{figure}

\begin{figure}
\begin{center}
\includegraphics[scale = 0.8]{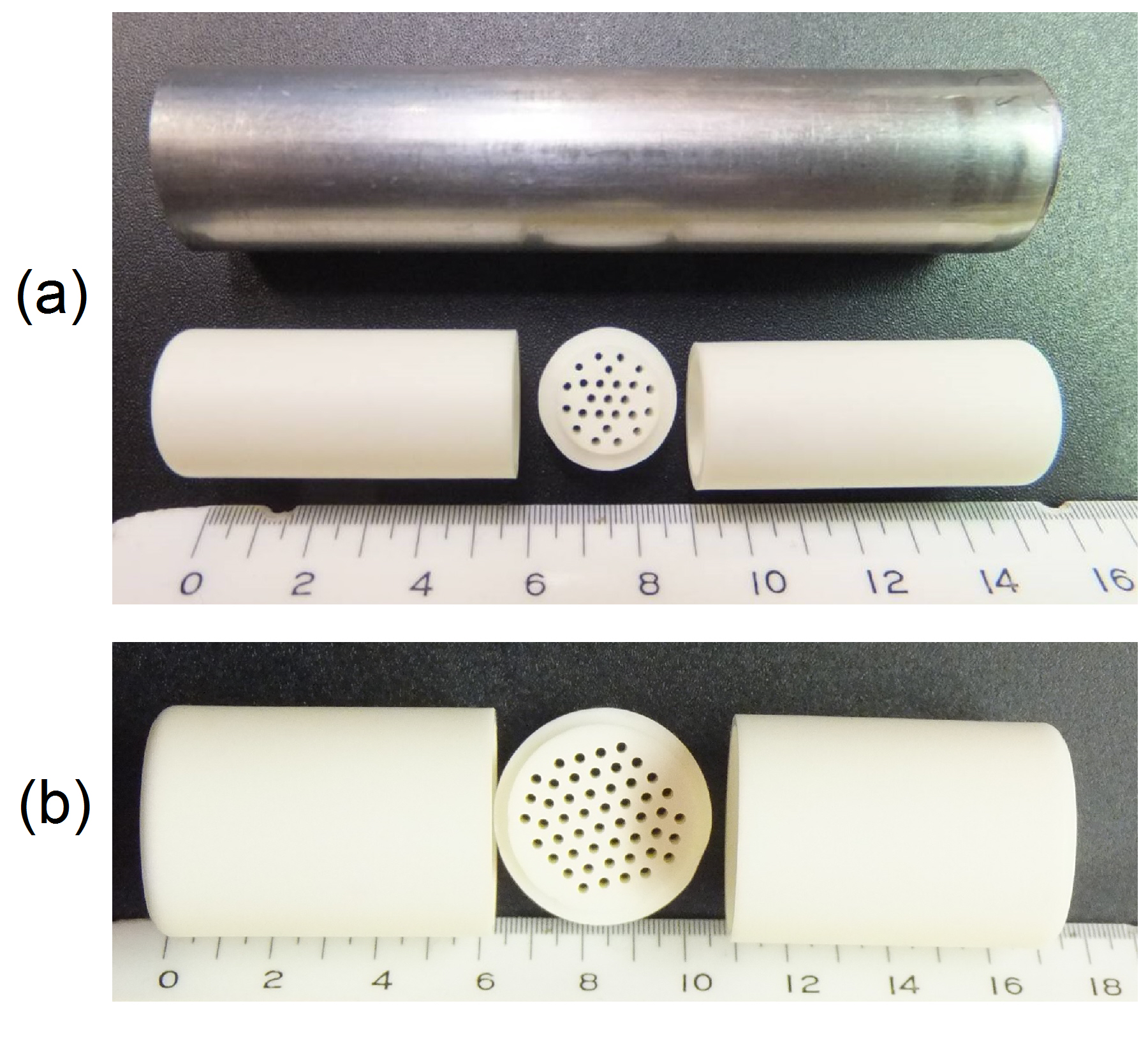}
\end{center}
\caption{(a) Slightly sub-2 ml crucible set that is designed to fit inside 11.9 mm I.D. Ta-tubing (shown above) and (b) 5-ml crucible set \cite{10}.}
\label{2}
\end{figure}

Figures \ref{1} and \ref{2} are photographs of three manifestations of the frit-disc crucible set that we have designed and used over the past few years\cite{10}. The 2 ml set (Fig.~\ref{1}) is the work-horse set for our group, with the 5 ml set (Fig.~\ref{2}b) being used for larger growths and the slightly sub-2 ml set (Fig.~\ref{2}a) being specially designed to be used inside 11.9 mm I.D. Ta-tubing. The frit-disc has a milled shoulder on each side to make what can be considered to be a “step-taper” that allows for simple and effective co-axial alignment of the three part assembly. When the crucibles and frit-disc are assembled, as in Fig.~\ref{1}b, the frit-disc extends into both growth and spin side crucibles. When the assembly is sealed into an amorphous silica tube, silica wool can be placed both below it, as needed, and above it, to act as a cushioning plug during the centrifugally enhanced decanting step (see reference \cite{3} for details and images). The silica wool holds the assemblage together during growth and the initial part of the decanting process.

The design requirement for this assemblage was that the cost of the frit-disc crucible set was less than the cost of three individual crucibles. By using the technically simpler milled shoulder or step-taper design there is no need to modify the two crucibles (i.e. no threading of the upper, inner portion of the crucibles) and the disc simply has square shoulders on both sides (again no threading). The frit-disc fits into the crucibles, but other than that, does not need to conform to tight specifications. As a result, the two crucibles are the same as normal 2 ml crucibles and the only cost requirement is that the frit-disc is less expensive than an individual crucible, a requirement that is readily met.

\section{Results}

The primary purpose of a frit-disc crucible set is the efficient separation of the crystalline solid phase from the remaining liquid phase in a manner that keeps both phases as clean as possible. This is a lovely and simple statement and, as such, actually encompasses several distinct variations on a theme. At its most basic level, the frit-disc removes silica wool from the growth volume and prevents it from contaminating the decanted liquid. This allows for quantitative analysis of the decanted liquid, as well as its reuse, as will be discussed in greater detail below. In many cases, the frit-disc also prevents (i) a subset of the grown crystals from either embedding themselves into the decanted liquid by being pushed through the silica wool filter and (ii) crystals from suffering from surface contamination of silica wool and residual liquid/oxide slag contamination that often would be trapped in the upper parts of the silica wool plug. It should be noted, though, that, in some cases, it is useful to have a physically softer filtering material. For example, in recent growths of CeAgSb$_{2}$ out of 5 ml crucibles\cite{4}, a single, very large crystal would grow and, in some cases would detach from the crucible wall and slam into the frit-disc and break into multiple pieces. For these growths we reverted back to the use of a silica wool plug to achieve a higher survival rate.

Another advantage the frit-disc has over a plug of silica wool is the relatively large hole size (on the order of 0.7 $\sim$ 1.0 mm). For many growths this is not an issue, and the size of the holes in the frit-disc is, if anything a mild source of worry in terms of losing some small amount of sub-mm sized crystals. For some cases though, especially at low temperatures, the surface tension of the remaining liquid becomes high enough that, even for accelerations of hundreds of g's, the crystals remain encased in a blob of molten solution that sits on top of the silica plug\cite{9}. This is what happens in the case of PtSn$_{4}$. Although PtSn$_{4}$ grows readily out of excess Sn and forms single crystal plates with dimensions of up to a cm on a side\cite{11}, it is exceptionally hard to separate from the remaining Sn when it is decanted at 350 $^{\circ}$C if silica wool is used. If, on the other hand, a frit-disc is used, then the excess Sn liquid does flow through the drilled holes. Again, it should be noted that in some cases the smaller number of bigger holes on a rigid disk can lead to problems that do not occur with plugs of silica wool. For example, if many plate like crystals cover all the holes, then, like leaves in a gutter, they can lead to decanted liquid over-flowing the frit and leaking out the side of the crucible. If this happens, then silica wool may be better for the decanting step. 

In some situations silica wool is simply not usable due to chemical attack from melt constituents (and/or their vapors). In such cases the fact that the frit disc is alumina can be important. A recent example of such a growth is the K$_{2}$Cr$_{3}$As$_{3}$ superconductor\cite{12}. For this growth the solution has to be segregated from any silica due to profound K-attack; Ta tubing is generally used to provide this segregation, but, Ta itself can be attacked by As as well as Cr. So alumina has to hold the liquid phase. Fig.~\ref{2}a shows a slightly sub-2 ml frit-disc crucible set that we had made\cite{10} to fit inside of our standard 11.9 mm I.D. Ta tubing. By welding the 3-piece frit-disc crucible set into a slightly longer Ta tube which was then sealed into an amorphous silica tube (to provide protection from oxidation)\cite{12}, single crystals of K$_{2}$Cr$_{3}$As$_{3}$ could be readily grown and separated from the excess solution.

Returning to the purity of the decanted liquid, the use of a frit-disc allows for the reuse and/or analysis of the liquid phase that is separated from the solid phase (if any). To make this point explicitly, if silica wool is used as a filter material, the decanted liquid is embedded in a significant percentage of the silica wool used. This means that during any attempt to reuse the decanted liquid there will be an abundance of silica wool included in the subsequent melt. Even if there is no chemical reaction between the silica wool and the melt, the thousands of silica fibers prevent the uniform growth of a single crystalline material.  

Being able to reuse clean, decanted solution presents a variety of economic and scientific opportunities. For new growths, the result of “total-spin”, i.e. the result of zero crystallization, is no longer so wasteful in terms of consumed elements. A “total-spin” is now a valuable data point in terms of the non-intersection of a liquidus surface and the spin side can simply be recycled, often for exploration of lower growth-temperature ranges. When the growth involves the use of precious-metal-rich solutions this is a very comforting safety net.  

Being able to reuse the spin, even after the growth of crystals can be very important as well. For example we recently have been working on very Cd-rich, rare earth cadmium binary compounds: $R$Cd$_{6}$ ($R$ = Y, Gd-Lu) crystalline approximant phases as well as a family of icosahedral, i-$R$Cd quasicrystals ($R$ = Y, Gd-Tm) with WDS determined stoichiometry ranging from GdCd$_{7.9}$ to TmCd$_{7.3}$\cite{13}. For both the approximant and quasicrystalline phases there was a need to perform neutron scattering measurements and this required the use of isotopically enriched Cd. The use of frit-disc crucibles was key to making the growth of $^{112}$Cd and $^{114}$Cd enriched samples possible.

\begin{figure}
\begin{center}
\includegraphics[scale = 0.5]{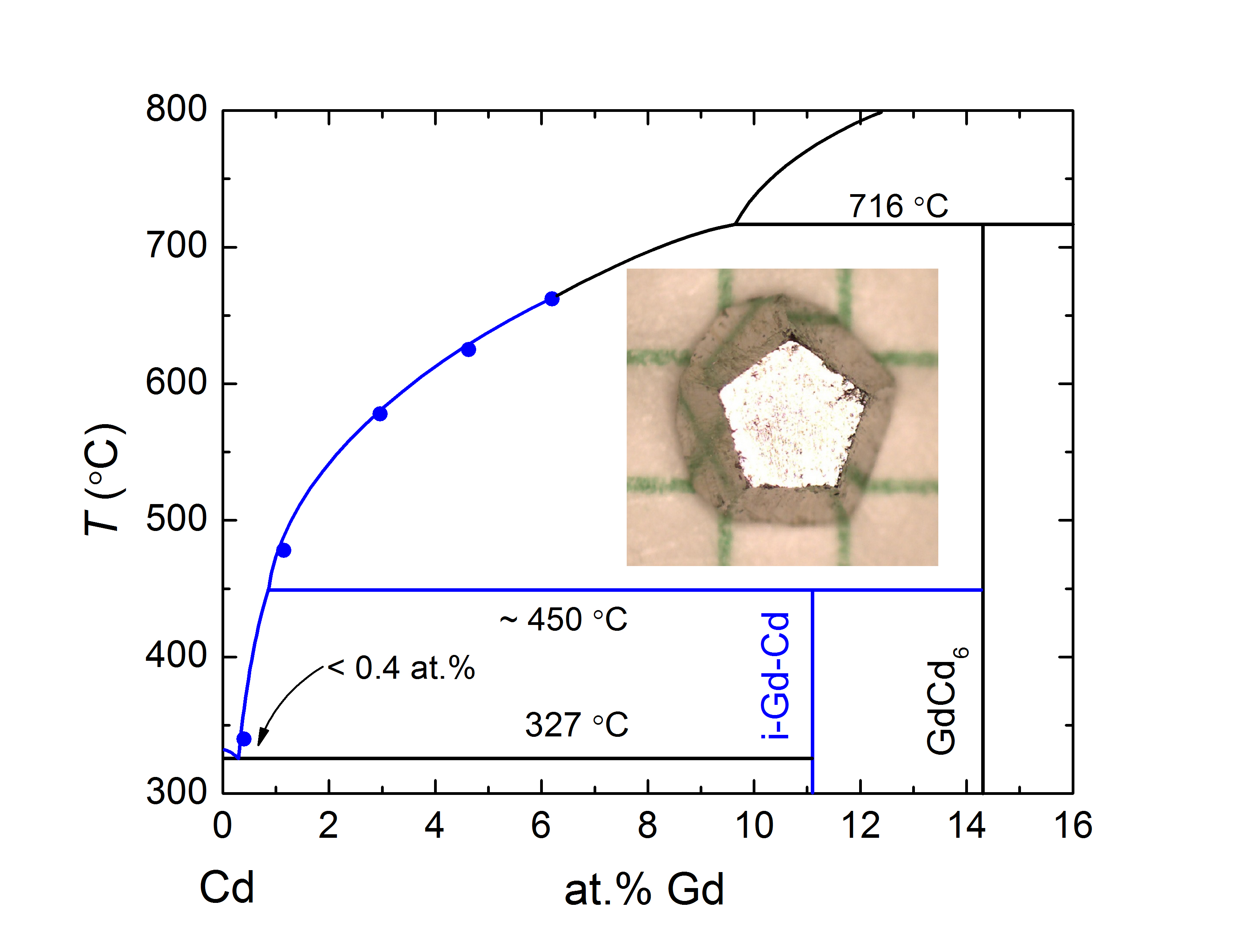}
\end{center}
\caption{Revised Gd-Cd binary phase diagram\cite{13,14}. Blue data points and new i-GdCd phase lines were obtained using frit-disc crucible sets. Picture in the center shows a single grain of i-GdCd quasicrystal.}
\label{3}
\end{figure}

Figure \ref{3} shows our newly determined, Cd-rich side of the Gd-Cd binary phase diagram\cite{13,14}. In order to grow the related i-TbCd, we needed to use a melt with a starting ratio of Tb:Cd = 0.8:99.2. As can be appreciated from the phase diagram, relatively little quasicrystalline phase can be grown during any one cooling cycle. Put in another way, if the decanted liquid were simply discarded, roughly 95$\%$ of the starting isotopically enriched Cd would be wasted. 

With the frit-disc crucible we were able to perform multiple growths of i-TbCd reusing the decanted solution from each prior growth. As an example, for an initial purchase of roughly 9.2 grams of $^{114}$Cd we split it into two, 5.0 and 4.2 gram amounts. Using the 5.0 grams amount we ran 3, sequential, growths of i-TbCd (with yields of 250 mg, 188 mg and 133 mg of i-TbCd, respectively). Using the 4.2 gram amount we ran two sequential growths of i-TbCd (with yields of 210 mg and 133 mg of i-TbCd respectively). Elemental Tb was added each time before reusing the decanted solution to reach desired Tb concentration. 

We then combined the remaining, decanted material and made one further i-TbCd growth with a yield of 206 mg. In total we were able to grow nearly 1.2 grams of large grains of i-TbCd (similar to, and even larger than, what is shown for i-GdCd crystal in Fig.~\ref{3})\cite{14}. 

At this point we used the remaining, decanted material to grow isotopically enriched TbCd$_{6}$. For this growth we used a starting ratio of Tb$_{7}$Cd$_{93}$. We made two, sequential growths, the first one yielding 2.7 grams of TbCd$_{6}$ and the second one yielding approximately 1 gram. By repeated reuse of the Cd-rich decanted solution we were able to double the amount of i-TbCd phase we could grow and use the remaining decanted solution to grow nearly 4 grams of TbCd$_{6}$.

Given that we have just been examining details of the re-determination of the $R$-Cd, binary phase diagram shown in Fig.~\ref{3}, it should be noted that this determination was made possible by the use of frit-disc crucible sets too. The clean separation of crystalline phase from remaining solution allowed us to determine the composition of the liquid at the decanting temperature. By assessing the mass of either RCd$_{6}$ or i-$R$Cd relative to the mass of decanted liquid and by varying decanting temperatures we were able to create the liquid-line\cite{13,14}. Although this might be considered to be a rather time consuming method of determining the liquidus line, it is one of the most accurate since solution growth is really the ultimate test of how phases solidify out of solution.

Another example of taking advantage of a reusable, clean decant can be found in our recent work in on Bi-Rh-S ternary phases\cite{15}. We were initially interested in studying single phase (and even single crystal) Bi$_{2}$Rh$_{3}$S$_{2}$, the S-analogue of Bi$_{2}$Rh$_{3}$Se$_{2}$, a compound that was reported to be a low-temperature superconductor with a higher temperature density-wave-like phase transition\cite{155}. Our initial growths had a melt stoichiometry of Rh$_{50}$Bi$_{25}$S$_{25}$ and were cooled from 1150 $^{\circ}$C to 800 $^{\circ}$C and produced crystalline material that, after careful analysis\cite{15}, turned out to be multi-phased consisting of Bi$_{2}$Rh$_{3}$S$_{2}$ and a new, Bi$_{2}$Rh$_{3.5}$S$_{2}$ phase. We were able to ultimately determine the temperature/composition ranges for primary solidification of the new, Bi$_{2}$Rh$_{3.5}$S$_{2}$ phase by repeated weighing of crystalline and decanted portions of the growth and reuse of the clean decant from each higher temperature growth for the next, lower temperature growth. To be specific, in this case, by starting with a Rh$_{50}$Bi$_{25}$S$_{25}$ composition, heating to 1150 $^{\circ}$C and decanting at 875 $^{\circ}$C, then reusing the decant, heating to slightly above 875 $^{\circ}$C and decanting at 825 $^{\circ}$C, then reusing the decant again, heating to slightly above 825 $^{\circ}$C and decanting at 800 $^{\circ}$C, we found that single phase Bi$_{2}$Rh$_{3.5}$S$_{2}$ can be grown from an initial melt stoichiometry of Rh$_{55}$Bi$_{22.5}$S$_{22.5}$ that is decanted at 775 $^{\circ}$C\cite{15}.

\begin{figure}
\begin{center}
\includegraphics[scale = 0.5]{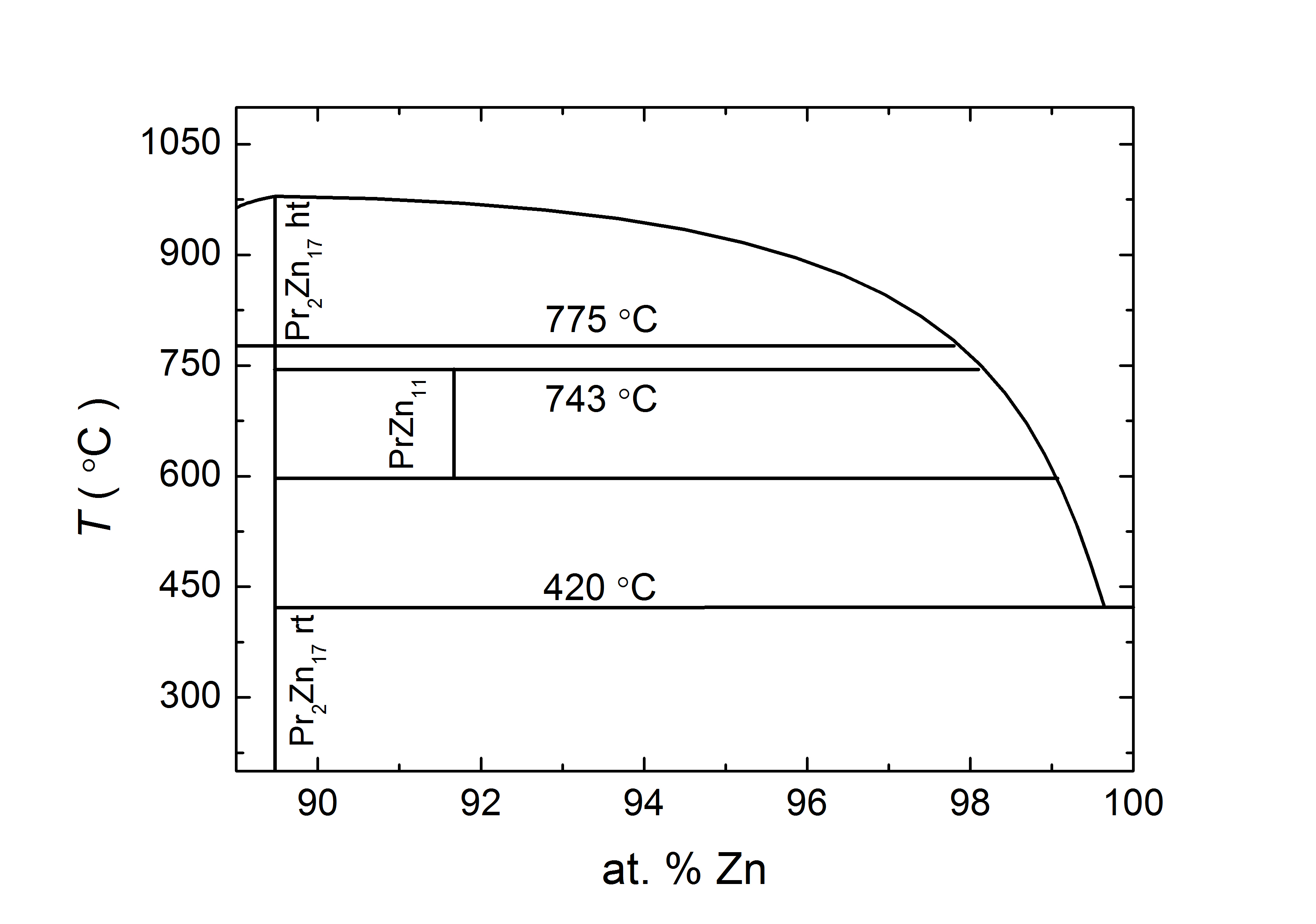}
\end{center}
\caption{Zn-rich part of the Pr-Zn binary phase diagram, redrawn from ref. \cite{16} based on our data. PrZn$_{11}$ appears to be a rare example of a binary line compound that decomposes catatectically. Below roughly 600 $^{\circ}$C, for solutions with roughly 0.005 at $\%$ Pr or less, the more Pr-rich Pr$_{2}$Zn$_{17}$ grows rather than PrZn$_{11}$.}
\label{4}
\end{figure}

In a similar manner we are currently exploring the possibility of a relatively rare example of a binary line compound, PrZn$_{11}$, having a catatectic decomposition in the region of 600 $^{\circ}$C. As part of a study of Pr-based compounds with Pr in a tetragonal point symmetry we grew single crystals of PrZn$_{11}$ out of excess Zn\cite{16}. Much to our surprise, when we cooled a melt with an initial composition of Pr$_{0.01}$Zn$_{0.99}$ from 900 $^{\circ}$C down to 500 $^{\circ}$C we ended up with PrZn$_{11}$ crystals in a distorted octahedral morphology $and$ hexagonal Pr$_{2}$Zn$_{17}$ that, based on morphology and location, appeared to have grown at lower temperatures. In order to confirm the order of crystallization we cooled a Pr$_{0.01}$Zn$_{0.99}$ melt from 900 $^{\circ}$C to 600 $^{\circ}$C, decanted and used the decant to perform a second growth from 800 $^{\circ}$C to 500 $^{\circ}$C. Whereas the first (900-600 $^{\circ}$C) growth gave single phase PrZn$_{11}$, the second (800-500 $^{\circ}$C) growth resulted in phase pure Pr$_{2}$Zn$_{17}$. To further underscore the fact that Pr$_{2}$Zn$_{17}$ grows at lower temperatures and for lower Pr-content in the melt, when a melt with an initial composition of Pr$_{0.005}$Zn$_{0.995}$ was cooled from 900 $^{\circ}$C to 500 $^{\circ}$C well-formed hexagonal plates of Pr$_{2}$Zn$_{17}$ grew. Fig.\ref{4} shows our current understanding of the Zn-rich part of the Pr-Zn binary phase diagram. Currently it appears that, like the case of SmAl$_{4}$\cite{17,18}, PrZn$_{11}$ grows out of excess Zn over a limited temperature range and then Pr$_{2}$Zn$_{17}$ growth reoccurs below the catatectic temperature. The fact that we can readily find PrAl$_{11}$ crystals in our growths, even though they have cooled to well below the catatectic temperature indicates that, like SmAl$_{4}$, the catatectic reaction is “sluggish”\cite{18}.

\section{Summary}

Frit-disc crucible sets, using a simple frit-disc that has a milled shoulder on each side to make what can be considered to be a “step-taper” that allows for simple and effective co-axial alignment of the three part assembly are, at this point, readily made using conventional Al$_{2}$O$_{3}$ processing techniques\cite{10}. Unlike the historically more usual plug of silica wool\cite{1,2,3}, the frit-disc allows for the clean separation of the liquid remaining at the end of a growth process(the decant) from any solid, ideally single crystalline, material. Such clean separation allows the decanted liquid to be reused and/or studied, as outlined in the examples given in this brief paper. Although there are a few situations or growths that might benefit from the use of a silica plug for filtering material (e.g. for cushioning of large or fragile crystals), the frit-disc should be the generic filtering device for most solution growths.

\section*{Acknowledgements}
The development of crystal growth techniques and tools is often evolutionary.  P.C.C would like to acknowledge and thank a string of former post-doctoral researchers who were part of this evolution:  Ian Fisher (Ta 3-cap)\cite{8}, Cedomir Petrovic (the initial BN frit-disc prototypes as well as the threaded frit-disc development\cite{9}) and Rongwei Hu (for having steered my attention toward LSP Ceramics\cite{10} for potential production). This work was supported by the U.S. Department of Energy, Office of Basic Energy Science, Division of Materials Sciences and Engineering. The research was performed at the Ames Laboratory. Ames Laboratory is operated for the U.S. Department of Energy by Iowa State University under Contract No. DE-AC02-07CH11358. N. H. Jo was supported by the Gordon and Betty Moore Foundation’s EPiQS Initiative through Grant GBMF4411

\end{document}